\documentclass[a4paper,12pt]{article}
\usepackage{amsmath}
\usepackage{amssymb}
\usepackage{latexsym}
\usepackage{graphicx}
\newcommand{\be}{\begin{eqnarray}}
\newcommand{\ee}{\end{eqnarray}}
\begin{document}
\begin{titlepage}

\begin{centering}
\vspace{.3in}
{\Large{\bf Energy Distribution in a Schwarzschild-like Spacetime}}
\\

\vspace{.5in} {\bf Th. Grammenos$^\dag$\footnote{thgramme@uth.gr}, I. Radinschi$^*$\footnote{radinschi@yahoo.com}},\\

\vspace{0.3in}
{\it $^\dag$  Department of Mechanical \& Industrial Engineering, University of Thessaly,\\
383 34 Volos, Greece\\
$^*$ Department of Physics, ``Gh. Asachi'' Technical University,\\
Iasi, 700050, Romania\\
}
\end{centering}

\begin{abstract}
\noindent In this paper, utilizing M\o ller's energy-momentum complex,
we explicitly evaluate the energy and momentum density associated with a metric describing a
four-dimensional, Schwarzschild-like, spacetime derived from an effective gravity coupled
with a U(1) gauge field in the context of a D3-brane dynamics in the classical regime, i.e., between
the asymptotic and the Planck regime.
\end{abstract}
\end{titlepage}

\section{Introduction}

The subject of the localization of energy still lacks an acceptable
answer and continues to be one of the most interesting and challenging
problems in General Relativity. For the solution of the problem many researchers
have computed the energy as well as the momentum and angular momentum associated
with various space-times. However, the different attempts of constructing an
energy-momentum density do not yield a generally accepted expression. After
Einstein first introduced energy-momentum complexes \cite{Einstein}, a plethora of
different energy-momentum complexes were
constructed, including those of Tolman \cite{Tolman}, Landau and Lifshitz \cite{Landau}, Papapetrou \cite{Papapetrou},
Bergmann and Thompson \cite{Bergmann}, Goldberg \cite{Goldberg}, Weinberg \cite{Weinberg} and M\o ller \cite{Moller}.
With the exception of M\o ller's approach, which could be utilized to any
coordinate system, the above energy-momentum complexes have a drawback, as they give
meaningful results only if the calculations are restricted to
quasi-Cartesian coordinates.

In 1973, Ch. Misner, K.Thorne and J.A.Wheeler sustained
that ``anybody who looks for a magic formula for local gravitational energy-momentum
is looking for the right answer to the wrong question'' \cite{Misner}. This is the actual
meaning of the nonuniqueness of the pseudotensor for the energy-momentum.
However, they concluded that the energy is indeed
localizable only for spherical systems. A few years later, Cooperstock and Sarracino \cite{Sarracino} demonstrated
that if the energy is localizable in spherical systems, then it is also localizable in any space-time.

In the 1990's H. Bondi sustained that ``In relativity a nonlocalizable form of energy is inadmissible,
because any form of energy contributes to gravitation
and so its location can in principle be found'' \cite{Bondi}. The idea of the energy-momentum complex
was severely criticized for a number of reasons, such as the nontensorial
nature of the energy-momentum complex and, hence, its dubious physical interpretation
\cite{Chandra}, and the fact that different energy distributions were
obtained by different energy-momentum complexes for the same geometry \cite{Bergqvist}.
Attempts to deal with the issue of the localization of the gravitational energy-momentum
include also the quasi-local approach \cite{Brown}.

The issue of the energy-momentum localization by use of the energy-momentum
complexes was revived by K.S. Virbhadra's pioneering work
\cite{Virbhadra1}.
In 1996, Aguirregabiria, Chamorro
and Virbhadra \cite{Virbhadra2} showed that four different energy-momentum complexes (ELLPW, standing for Einstein,
Landau-Lifshitz, Papapetrou, Weinberg) yield the same
energy distribution for any non-static, spherically symmetric metric of the Kerr-Schild class.
Furthermore, their results complied with the quasi-local mass definition given earlier by Penrose
and Tod \cite{Brown}. In 1999, Chang, Nester and Chen \cite{Chang} proved that every energy-momentum
complex is associated with a legitimate Hamiltonian boundary term, thus
supporting the quasi-locality of energy-momentum complexes and, hence, their
acceptance.

The large number of interesting results
recently obtained by many researchers point out that the energy-momentum
complexes are powerful tools for evaluating the energy and momentum in a
given space-time \cite{Chamorro}. Important works are done with the
energy-momentum complexes in 2- and 3-dimensional space-times \cite{Vagenas}. Also, we
point out some interesting papers \cite{Korunur} which demonstrate that the (ELLPW)-
and Bergmann prescriptions yield the same results as their tele-parallel
gravity versions for a given spacetime.

In this work we have chosen M\o ller's prescription, because it is not restricted
to quasi-Cartesian coordinates, as pointed out earlier. Furthermore, there are many results \cite{Yang} that
recommend this prescription for the localization of energy.
Thus, we implement the M\o ller
prescription and calculate the energy density for a metric describing a
Schwarzschild-like geometry in the classical regime, in the context of
a D3-brane dynamics study. The calculations are performed with Mathematica and Maple, the latter having
attached the GrTensor platform. Throughout the paper we used geometrized units ($G=1$, $c=1$)
and let Greek indices run from $0$ to $3$. The remainder of the paper
is organized as follows. In Sec.2 we present the Schwarzschild-like
spacetime, while in Sec.3 we give a description of M\o ller's
prescription. In Sec.4 we explicitly determine the energy and momentum
distributions in the Schwarzschild-like spacetime using M\o ller's
prescription. Finally, Sec.5 is devoted to a summary of the obtained results
and concluding remarks.

\section{The Schwarzschild-like Geometry}

In a recent work, S. Kar and S. Majumdar \cite{Kar} considered the evolution of
gravity on a D3-brane in a noncommutative string theory. In particular,
the authors relied on the fact, that a D3-brane world-volume incorporating
Einstein's gravity coupled to the nonlinear theory of Maxwell may provide a
framework for the formulation of an effective theory of quantum gravity at
the Planck scale. In going towards the Planck regime, they combined a theory
of effective gravity with a U(1) gauge field thus obtaining, in the classical
regime, a Schwarzschild-like and a Reissner-Nordstr\o m-like
solution in (3+1) dimensions. In this work, we focus on the Schwarzschild-like solution.

Specifically, the authors considered a Euclidean world-volume spanned by\break $(y_1,y_2,y_3,y_4)$ with a signature $(+,+,+,+)$, where
the Minkowski signature may be obtained by  analytic continuation $y_4 \rightarrow it$.
Starting from the asymptotic regime with a flat D3-brane one can generalize the
brane's description by including a slow variation in the induced metric $g_{\mu\nu}$ in going into the
classical regime. The dynamics of the brane is then governed by the coupling of
the general-relativistic action with an appropriate Dirac-Born-Infeld action, whereby
the authors considered a static gauge condition on spacetime. Thus, the complete action becomes
\begin{equation}\label{1}
S=\frac{1}{16\pi}\int d^4 y \sqrt{g} R+S_{\text{DBI}}
\end{equation}
with $R$ the scalar curvature, and $S_{\text{DBI}}$ the Dirac-Born-Infeld action. After
expanding, one obtains
\begin{equation}\label{2}
S=\int d^4 y \sqrt{g} \left(\frac{1}{16\pi}R-
\frac{1}{4}g^{\mu\nu}g^{\lambda\rho}{\cal F}_{\mu\lambda}{\cal F}_{\nu\rho}+{\cal O}({\cal F}^4)+\ldots\right)
\end{equation}
where ${\cal F}_{\mu\nu}$ is the U(1) gauge field and higher order terms may be neglected in the classical
regime. For the gauge invariant field strength we have
\begin{equation}\label{3}
\bar{{\cal F}}_{\mu\nu}=({\cal B}+2\pi \alpha^{\prime}F)_{\mu\nu}
\end{equation}
$\alpha^{\prime}$ denoting the slope parameter in the open bosonic string theory and
$F_{\mu\nu}$ the electromagnetic field tensor.
The equation of motion for the gauge field is
\begin{equation}\label{4}
\partial_{\mu}{\cal F}^{\mu\nu}=0.
\end{equation}
The field equations obtained by the variation of the action (1) are
\begin{equation}\label{5}
R_{\mu\nu}-\frac{1}{2}R g_{\mu\nu}=8\pi T_{\mu\nu}
\end{equation}
where the energy-momentum tensor $T_{\mu\nu}$ is
\begin{equation}\label{6}
T_{\mu\nu}=\frac{1}{\sqrt{g}}\frac{\delta S_{\text{DBI}}}{\delta g^{\mu\nu}}=\frac{1}{2}\left(\frac{1}{4}
g_{\mu\nu}{\cal F}_{\mu^{\prime}\nu^{\prime}}{\cal F}^{\mu^{\prime}\nu^{\prime}}-{\cal F}_{\mu\lambda}{\cal F}^{\lambda}_{\nu}\right).
\end{equation}
The uniform electromagnetic field on the brane is expressed by its components\break
$\mathbf{E} = (0, E_{2}, E_{3})$ and $\mathbf{B} = (0, B_{2}, B_{3})$. The U(1) gauge potential is
\begin{equation}\label{7}
A_{\mu}=\left(\frac{-iQ_e}{r}, 0, 0, Q_m \cos\theta\right)
\end{equation}
with $Q_{e}$, $Q_{m}$ constants denoting the electric and magnetic charge, respectively, while
the electromagnetic field takes the (anti-parallel configuration) form
\begin{equation}\label{8}
\mathbf{E}=-\frac{Q_e}{r^2}\hat{r} \quad\quad \text{and} \quad\quad \mathbf{B}=\frac{Q_m}{r^2}\hat{r}.
\end{equation}
The effective metric on the brane is given by \cite{Seiberg}
\begin{equation}\label{9}
G_{\mu\nu}=g_{\mu\nu}-({\cal B}g^{-1}{\cal B})_{\mu\nu}
\end{equation}
with ${\cal B}_{\mu\nu}$ a constant 2-form induced on the world volume of the D3-brane.\footnote{In the
presence of a D-brane, a constant ${\cal B}$-field cannot be gauged away and can be reinterpreted as a constant
magnetic field on the brane.}
In the classical regime, for ${\cal B}=0$, the action (\ref{2}) reduces to that of General Relativity coupled
to Maxwell's electromagnetism, while (\ref{3}) becomes $\bar{{\cal F}}_{\mu\nu}=2\pi \alpha^{\prime}F_{\mu\nu}$. Then, the effective metric on the brane reads
\begin{equation}\label{10}
G_{\mu\nu}=g_{\mu\nu}-(\bar{{\cal F}}g^{-1}\bar{{\cal F}})_{\mu\nu}+{\cal O}({\cal F}^4)+\ldots
\end{equation}
Since $T_{\mu\nu}$ is weak, the gravitational solution can be approximated
by the Schwarzschild geometry. Thus, a (semi)classical
solution of equations (\ref{4}) and (\ref{5}) can be obtained and, ignoring higher order terms in (\ref{10}) as in (\ref{2}),
the line element for the effective metric finally becomes
\begin{equation}\label{11}
\begin{split}
ds^{2} = &-\left(1-\frac{2M}{r}\right)\left(1-\frac{Q_e^2}{r^4}\right)dt^{2}
+\left(1-\frac{2M}{r}\right)^{-1}\left(1-\frac{Q_e^2}{r^4}\right)^{-1}dr^2\\
&+\left(1-\frac{Q_{m}^{2}}{r^{4}}\right)r^{2}\,d\theta ^{2}
+\left(1-\frac{Q_{m}^{2}}{r^{4}}\right)^{-1}r^{2}\,\sin^{2}\theta \,d\varphi ^{2}.
\end{split}
\end{equation}
Actually, the metric given in \cite{Kar} has a positive sign in front of the first term. However, the negative sign here arises from the
transformation $t \rightarrow it$ in going from the Euclidean to the Lorentzian signature. The above line element describes an, asymptotically flat,
Schwarzschild-like geometry, which becomes the known Schwarzschild solution when $Q_{e}=Q_{m}=0$. This (semi-)classical solution
is of Petrov type I and it is not spherically symmetric. It becomes of Petrov type D and acquires spherical symmetry only
when $Q_{m}=0$. By computing the Kretschmann scalar, one
can see that, in the general case, i.e., when the electric as well as the magnetic charge are nonzero, the above solution has two curvature singularities
(at $r=0$ and at $r=\sqrt{Q_{m}}$). Furthermore, there are three horizons, namely at $r=2M$, at $r=\sqrt{Q_{e}}$, and at $r=\sqrt{Q_{m}}$).
In fact, one of the curvature singularities turns out to be also a horizon. As it is evident, the above solution describes
a highly exotic situation, which should be studied further and in more detail.

\section{M\o ller's Prescription}

In the Introduction we have pointed out the importance of the energy-momentum complexes
for the energy-momentum localization thereby stressing the role of
the M\o ller prescription in this context.
The M\o ller energy-momentum complex is an efficient tool for the energy-momentum
localization and allows obtaining satisfactory results for the energy and momentum
distributions in the case of a general-relativistic system.

The M\o ller energy-momentum complex in a four-dimensional background \cite{Moller} is given as
\begin{equation}\label{12}
\mathcal{J}_{\nu }^{\mu }=\frac{1}{8\pi }\xi _{\nu \,\,,\,\lambda }^{\mu
\lambda }
\end{equation}
where M{\o }ller's superpotential $\xi _{\nu }^{\mu \lambda }$ is of the form
\begin{equation}\label{13}
\xi _{\nu }^{\mu \lambda }=\sqrt{-g}\left( \frac{\partial g_{\nu \sigma }}{%
\partial x^{\kappa }}-\frac{\partial g_{\nu \kappa }}{\partial x^{\sigma }}%
\right) g^{\mu \kappa }g^{\lambda \sigma }
\end{equation}
with the antisymmetric property
\begin{equation}\label{14}
\xi _{\nu }^{\mu \lambda }=-\xi _{\nu }^{\lambda \mu }.
\end{equation}

It is easily seen that M\o ller's energy-momentum complex satisfies the
local conservation equation
\begin{equation}\label{15}
\frac{\partial \mathcal{J}_{\nu }^{\mu }}{\partial x^{\mu }}=0
\end{equation}
where $\mathcal{J}_{0}^{0}$ is the energy density and $\mathcal{J}_{i}^{0}$,
$i = 1,2,3$, are the momentum density components.

Thus, in M\o ller's prescription the energy and momentum for a
four-dimensional background are given by
\begin{equation}\label{16}
P_{\nu }=\int \int \int \mathcal{J}_{\nu }^{0}dx^{1}dx^{2}dx^{3}.
\end{equation}
Specifically, the energy of a physical system in a four-dimensional
background is
\begin{equation}\label{17}
E=\int \int \int \mathcal{J}_{0}^{0}dx^{1}dx^{2}dx^{3}.
\end{equation}
In this prescription the calculations are not anymore restricted to quasi-Cartesian coordinates.
They can be utilized in any coordinate system.

\section{Energy and Momentum Density Distributions}

First, we have to evaluate the superpotentials. There are eight nonzero M\o ller
superpotentials:
\begin{equation}\label{18}
\xi^{01}_{0}=-\xi^{10}_{0}=-\left[4Q_e^2\frac{1}{r^3}\left(1-\frac{2M}{r}\right)+2M\left(1-\frac{Q_e^2}{r^4}\right)\right]\sin\theta
\end{equation}
\\
\begin{equation}\label{19}
\xi^{12}_{2}=-\xi^{21}_{2} =
-\frac{2}{r^3}\left(1-\frac{2M}{r}\right)\left(\frac{Q_m^2+r^4}{Q_m^2-r^4}\right)(r^4-Q_e^2)\sin\theta
\end{equation}
\\
\begin{equation}\label{20}
\xi^{13}_{3}=-\xi^{31}_{3}=2r\left(1-\frac{2M}{r}\right)\left(1-\frac{Q_e^2}{r^4}\right)\left(\frac{3Q_m^2-r^4}{Q_m^2-r^4}\right)\sin\theta
\end{equation}
\\
\begin{equation}\label{21}
\xi^{23}_{3}=-\xi^{32}_{3}=-\frac{2r^4 \cos\theta}{Q^2_m-r^4}.
\end{equation}
\\
\noindent By substituting the superpotentials given by (\ref{18}-\ref{21}) into (\ref{12}) we get for the
energy density distribution
\begin{equation}\label{22}
J^{0}_{0}=\frac{Q_{e}^{2}(3r-10M)\sin\theta}{2\pi r^5}.
\end{equation}
Furthermore, it is found out that all the momentum density distributions vanish. Now, substituting (\ref{22}) into (\ref{17})
and evaluating the integral, we obtain the energy contained in a ``sphere" of radius R:
\begin{equation}\label{23}
E(R)=M-2Q_e^2 \left(-\frac{5M}{2R^4}+\frac{1}{R^3}\right).
\end{equation}
This result, depending only on the electric charge, gives the effective gravitational mass for the spacetime considered. At very large distances,
i.e. at the asymptotic limit, the energy equals the (ADM) mass $M$. However, the energy equals the mass M also
for the finite radius $R=\frac{5}{2} M$, an unexpected and remarkable result for which, beyond speculations, no reasonable explanation has be found.
At this stage of the investigation, one can only conjecture that this result would be attributed to the higly exotic character
of the object having the spacetime described by eq.(11).

\section{Discussion}

A D3-brane in the presence of a uniform electromagnetic field in an open bosonic string theory is
considered. By including a slow variation in the induced metric tensor $g_{\mu\nu}$ in the classical regime, one ends up
with an action describing the D3-brane dynamics and consisting in the coupling of the Einstein-Hilbert
action to a Dirac-Born-Infeld action, whereby a static gauge condition is assumed. The aforementioned complete
action leads, in the classical regime, to an effective metric describing the geometry of a Schwarzschild-like
spacetime on the D3-brane \cite{Kar} with curious properties that need to be further investigated.

In this work, we have explicitly calculated the energy and momentum densities for this effective metric.
The geometry considered is spherically symmetric when the magnetic charge $Q_{m}=0$. Furthermore, if the electric as well as the
magnetic charge vanishes, the geometry is identical with that describing the spacetime exterior to a Schwarzschild black hole.
The energy and momentum densities are computed using the M{\o }ller energy-momentum complex.
It is found that all the momentum densities vanish, while the effective gravitational mass, i.e. the total energy contained in a
"sphere" of radius R in the considered, Schwarzschild-like, spacetime, depends on the mass $M$ and
the electric charge $Q_{e}$. At the asymptotic limit, the energy is equal
to the (ADM) mass $M$. However, this value is also obtained for a finite radius R, a result that remains an open question
to be answered. Last but not least, the paper sustains Lessner's argumentation \cite{Lessner}
supporting M{\o }ller's prescription as a powerful tool for describing the concepts of energy and momentum in
General Relativity.

Work on the computation of the energy and momentum distributions for the\break Schwarzschild-like geometry
generalized to a black hole solution on a noncommutative D3-brane in a static gauge at the Planck
scale, is in progress.

\section*{Acknowledgements}
The authors are indebted to Dr. E.C. Vagenas for stimulating discussions and useful suggestions
and to Dr. G.O. Papadopoulos for valuable comments. They, also, thank the unknown referees for their
suggestions which helped correcting and improving the final form of the manuscript.

\end{document}